\documentclass[11pt,prl,twocolumn,english,aps,reprint,superscriptaddress,notitlepage,nobibnote,preprintnumbers,showpacs]{revtex4-2}

\usepackage{amsfonts}
\usepackage{bbm,bm}
\usepackage[breaklinks,urlbordercolor={1 1 1}]{hyperref}
\usepackage{graphicx}
\usepackage{amsmath}
\usepackage{amssymb}
\usepackage{mathtools}
\usepackage{hyperref}
\usepackage{graphicx}
\usepackage{color}
\usepackage{comment}
\usepackage{subfigure}
\usepackage{dcolumn}
\usepackage[utf8]{inputenc} 
\usepackage{multirow}
\usepackage{float}
\usepackage{booktabs}

\usepackage{ulem,fancyvrb}
\usepackage{xcolor}

\newcommand{\be}{\begin{equation}} 
\newcommand{\ee}{\end{equation}}

\newcommand{\ba}{\begin{eqnarray}}
\newcommand{\ea}{\end{eqnarray}}

\newcommand{\yp}{{Y\rm_P}}
\newcommand{\ev}{{ \rm eV}}

\newcommand{\neff}{N_{\rm eff}}
\newcommand{\xin}{\xi_{\nu}}

\begin{document}

\newcommand{\ITP}{\affiliation{CAS Key Laboratory of Theoretical Physics, Institute of Theoretical Physics,\\
Chinese Academy of Sciences, Beijing 100190, China}}

\newcommand{\HIAS}{\affiliation{School of Fundamental Physics and Mathematical Sciences, Hangzhou Institute for Advanced
Study, UCAS, Hangzhou 310024, China}}

\newcommand{\PKU}{\affiliation{Center for High Energy Physics, Peking University, Beijing 100871, China}}

\newcommand{\UCAS}{\affiliation{School of Physical Sciences, University of Chinese Academy of Sciences, Beijing 100049, P.\ R.\ Chin}}



\title{Neutrinophilic $\mathbf{\Lambda}$CDM Extension for EMPRESS, DESI and Hubble Tension}


\author{Yuan-Zhen Li} 
\email{liyuanzhen@itp.ac.cn} 
\affiliation{
CAS Key Laboratory of Theoretical Physics, Institute of Theoretical Physics, \\ Chinese Academy of Sciences, Beijing 100190, P.\ R.\ China}
\affiliation{
School of Physical Sciences, University of Chinese Academy of Sciences, Beijing 100049, P.\ R.\ China}

\author{Jiang-Hao Yu}
\email{jhyu@itp.ac.cn}
\affiliation{
CAS Key Laboratory of Theoretical Physics, Institute of Theoretical Physics, \\ Chinese Academy of Sciences, Beijing 100190, P.\ R.\ China}
\affiliation{
School of Physical Sciences, University of Chinese Academy of Sciences, Beijing 100049, P.\ R.\ China}
\affiliation{
School of Fundamental Physics and Mathematical Sciences, Hangzhou Institute for Advanced
Study, UCAS, Hangzhou 310024, China}

\begin{abstract}
A number of recent cosmological observations have indicated the presence of new physics beyond the $\Lambda$CDM model. 
Combining observations from EMPRESS on helium abundance and DESI on baryon acoustic oscillations with Hubble tension, we show that all of them can be explained concurrently with a extension of the $\Lambda$CDM model with primordial neutrino asymmetry $\xin$ and additional contribution to the effective number of neutrinos $\delta\neff$.
Based on the accurate treatments of neutrino decoupling and BBN processes, we present state-of-the-art constraints on neutrino asymmetry for the fixed or varying $\neff$.
Comparing different extensions of the $\Lambda$CDM model, we show that the neutrinophilic $\Lambda$CDM extension with $\xin = 0.056 \pm 0.017 $ and $\delta\neff = 0.41 \pm 0.16$ is preferred by current observations, while the Hubble tension in this model is also alleviated to be $2.2 \sigma$. 
\end{abstract}

\maketitle


\textbf{\textit{  Introduction.}} We are now in the era of precision cosmology, where observations of both the early and late universe provide key insights into beyond the Standard Model (SM) physics. Precision measurements of the Cosmic Microwave Background (CMB) over the past decade by Planck~\cite{Planck:2018vyg}, ACT~\cite{ACT:2020gnv,ACT:2023kun}, and SPT~\cite{SPT-3G:2021eoc} are consistent with the $\Lambda$ Cold Dark Matter ($\Lambda$CDM) model, which incorporates non-baryonic cold dark matter and dark energy as a cosmological constant, $\Lambda$~\cite{Baumann:2015rya,Komatsu:2022nvu,Chang:2022tzj}. However, recent data indicates possibility of new physics beyond the $\Lambda$CDM framework.

For late universe, the most notable indication is the "Hubble tension", where measurements of the Hubble constant from local distance ladders (using Cepheids and other anchors) differ from those obtained via CMB observations based on the $\Lambda$CDM model, at the $ 4-5 \sigma$ level~\cite{Planck:2018vyg,eBOSS:2020yzd,Riess:2016jrr,Riess:2018uxu,Pesce:2020xfe,Riess:2021jrx}. This significant discrepancy between early and late universe observations suggests potential shortcomings in the $\Lambda$CDM model~\footnote{Specifically, the SH0ES Cepheid-based distance ladder gives $H_0 = 73.04 \pm 1.04 \, \rm km \, s^{-1}Mpc^{-1}$~\cite{Riess:2021jrx}, while Planck reports $H_0 = 67.36 \pm 0.54 \, \rm km \, s^{-1}Mpc^{-1}$ based on the $\Lambda$CDM model~\cite{Planck:2018vyg}.}, leading to increased interest in recent years, see~\cite{Verde:2019ivm,DiValentino:2020zio,DiValentino:2021izs,Kamionkowski:2022pkx} for recent reviews on possible solutions. 
Moreover, observations from the Dark Energy Spectroscopic Instrument (DESI) collaboration also hints on possible cosmological scenarios beyond $\Lambda$CDM~\cite{DESI:2024mwx}. DESI measures the imprint of the sound horizon at the drag epoch, $r_d$, on galaxy, quasar, and Lyman $\alpha$ forest clustering, providing the strongest constraints to date on both the expansion history and growth rate of the Large Scale Structure (LSS). When combining DESI BAO data with CMB~\cite{Planck:2018vyg,ACT:2023kun} and PathenonPlus~\cite{Scolnic:2021amr} (DESY5~\cite{Abbott:2024agi}) data, the DESI collaboration finds that the dynamical dark energy model is favored over the $\Lambda$CDM model at the 2.5$\sigma$ (3.9$\sigma$) significance level.


On the other hand, the recent EMPRESS survey of primordial elements from Big Bang Nucleosynthesis (BBN) in extremely metal-poor galaxies reports a primordial helium-4 determination of $Y_P = 0.2370^{+0.0034}_{-0.0033}$~\cite{empress}, about 3$\sigma$ deviation than 
the SM prediction~\cite{Pitrou:2018cgg}. Since proton-neutron decoupling at the onset of BBN is influenced by electron neutrino properties, the EMPRESS result suggests a non-zero Primordial Neutrino Asymmetry (PNA) $\xi_{\nu}$~\cite{empress,Burns:2022hkq,Escudero:2022okz,Froustey:2024mgf}.
However, previous analyses overlook several aspects: first, the effects of PNA on the neutrino decoupling process; second, the radiative and nucleon mass corrections to weak rates for accurate BBN predictions; third, the constraining power of BAO observations on PNA, although they are widely used to constrain neutrino properties like the sum of neutrino masses $\sum m_\nu$ and the effective number of neutrinos $\neff$. 

In this letter, we present the first detailed analysis on the implications of PNA by combining the EMPRESS BBN, CMB and DESI BAO observations, based on the accurate BBN predictions from the companion paper~\cite{Li:2024gzf}, which addresses the above three aspects with only the BOSS BAO data~\cite{Beutler:2011hx,Ross:2014qpa,BOSS:2016wmc}. For both fixed and freely-varying $\neff$, we present state-of-the-art constraints on PNA and discuss the implications of including LSS data. Surprisingly, we find that both the peculiar EMPRESS BBN and DESI BAO observations can be simultaneously explained by a simple neutrinophilic $\Lambda$CDM model with PNA $\xi_\nu$ and additional contribution to effective number of neutrinos $\delta \neff$, while alleviating the Hubble tension to the 2.2$\sigma$ level, although alternative models can explain each observation individually.
This neutrinophilic $\Lambda$CDM model is also preferred over the $\omega_0 \omega_a$CDM model and other $\Lambda$CDM extensions.


\textbf{\textit{  Primordial Neutrino Asymmetry and implications.}} Primordial neutrino asymmetry is usually parameterised with the degeneracy parameters $\xi_{\nu_\alpha} \equiv \mu_{\nu_\alpha}/T_{\nu_\alpha}$~\footnote{For sufficient large $\xi_{\nu_\alpha}$, the total lepton asymmetry is also dominated by the neutrino asymmetry, $\eta_L \simeq \eta_\nu$, due to the electric charge neutrality of the early universe.}, defined as the chemical potential for the flavour $\alpha$ neutrino normalised to its temperature, so that
\begin{equation}
    \label{eq:etaL}
    \eta_\nu \equiv \frac{1}{n_{\gamma}}\sum_{\alpha=e,\mu,\tau}(n_{\nu_\alpha}-n_{\bar{\nu}_\alpha}) \simeq \frac{\pi^2}{33 \zeta(3)} \sum_{\alpha=e,\mu,\tau} \xi_{\nu_\alpha} \ ,
\end{equation}
where $n_{\gamma}$ represents the photon number density, and $n_{\nu_{\alpha}} (n_{\bar{\nu}_{\alpha}})$ refers to the (anti)neutrino number density for flavour $\alpha$. In Eq.~\ref{eq:etaL}, the SM value for the neutrino-photon temperature ratio $T_{\nu_{i}}/T_{\gamma} = (4/11)^{1/3}$ and a small degeneracy $\xi_{\nu_\alpha} \ll 1$ are assumed, such that higher-order terms in $\xi_{\nu_\alpha}$ can be neglected. 
Despite the sphaleron process in the early universe~\cite{Kuzmin:1985mm,Khlebnikov:1988sr,Harvey:1990qw,Dreiner:1992vm}, various models can generate a PNA $\eta_\nu$ that is much larger than the baryon asymmetry of the universe $\eta_B \equiv n_B/n_{\gamma} \sim \mathcal{O}(10^{-10})$~\cite{Casas:1997gx,Dolgov:1989us,Bajc:1997ky,Asaka:2005pn,Asaka:2005an,Pilaftsis:2003gt,Borah:2022uos,Kawasaki:2002hq,Kawasaki:2022hvx}. 

During evolution of the universe, PNA induces significant changes in various epochs, including neutrino decoupling, BBN, CMB, and LSS formation. In the following, we will explore these effects in detail based on an accurate treatment of full neutrino transport during decoupling in the companion paper~\cite{Li:2024gzf}. 


For neutrino decoupling, both $\neff$ and the neutrino spectral distortions are modified in the presence of PNA.
In the flavour-equilibrated case, where $\xi_{\nu_e} = \xi_{\nu_\mu} = \xi_{\nu_\tau} = \xin$, and accounting for flavour oscillations, matter effects, finite-temperature QED (FTQED) corrections up to order $\mathcal{O}(e^3)$, and full neutrino-electron and neutrino-neutrino collision terms, we find that
\begin{equation}\label{eq:neff}
    \neff = \neff^{\rm SM} + 3 \left( \frac{30}{7 \pi^2} \xi_{\nu_\alpha}^2 + \frac{15}{7 \pi^4} \xi_{\nu_\alpha}^4 \right) + 0.0102 \, \xin^2 \, ,
\end{equation}
where $\neff^{\rm SM} = 3.0440$ is the resulting $\neff$ for the SM case~\cite{Froustey:2020mcq,Bennett:2020zkv,Akita:2020szl}, and the second (third) term represent the contribution from PNA in (beyond) the instantaneous decoupling limit. In addition, neutrinos and antineutrinos experience different dynamics in the presence of PNA, leading to different spectral distortions with respect to corresponding thermal distributions.


In the BBN epoch, PNA primarily affects the neutron-proton conversion process and the final neutron-proton ratio after decoupling. Including these, along with corrections to weak rates for neutron-proton conversion (such as zero and finite temperature radiative corrections, nucleon mass corrections, and spectral distortions of neutrinos and antineutrinos), we provide state-of-the-art predictions for BBN in~\cite{Li:2024gzf}, which will be used in the Markov Chain Monte Carlo (MCMC) analysis.

Since neutrinos decouple completely during the formation of the CMB and LSS, the effects of the PNA are primarily expressed through two indirect effects: (1) the altered helium abundance $ Y_P$ from BBN, and (2) changes in the expansion history due to the increased $\neff $ and the non-relativistic effects of massive neutrinos.
For the CMB, the dominant effect arises from the first term, which modifies the tail of the CMB angular power spectrum through diffusion damping. The second term influences the matter-radiation equality, the last scattering epoch, and the BAO scale $r_d $, leaving imprints on the matter power spectrum and galaxy distributions~\footnote{Additionally, the free-streaming effect of massive neutrinos is alo enhanced due to the increased neutrino momentum from the PNA, resulting in greater suppression of the matter power spectrum at small scales. However, this effect is nearly negligible compared to the impact of neutrino masses. See~\cite{Li:2024gzf} for more details.}.

In summary, the neutrino cosmology with non-zero PNA $\xi_\nu$ and $\delta \neff$ would modify the whole cosmological history of the $\Lambda$CDM model. In the following, we call the modified cosmology due to extensions of neutrino properties as "the neutrinophilic $\Lambda$CDM model". 


\textbf{\textit{  Methodology.}} We will utilize the following datasets for our cosmological analysis:
\begin{description}
\item[BBN] For helium abundance, we use the EMPRESS result $\yp = 0.2370^{+0.0034}_{-0.0033}$~\cite{empress}. For deuterium abundance, we adopt the Particle Data Group (PDG)'s recommended value $10^5 \times {\rm D/H} = 2.547 \pm 0.025$~\cite{pdg}. 

\item[CMB] We use data from Planck collaboration, including low- and high-$\ell$ TT, TE, EE spectra, and the reconstructed CMB lensing spectrum~\cite{Planck:2018vyg,Planck:2018lbu,Planck:2019nip}.

\item[BAO] We adopt the latest DESI BAO measurements, which probe the sound horizon imprints at the drag epoch $r_d$ in the LSS~\cite{DESI:2024mwx}. 
These measurements directly provide constraints on the transverse comoving distance $D_M(z)$ and the equivalent distance $D_H(z)$, or their combination $D_V(z)/r_d = \left(z D_M(z)^2 D_H(z)\right)^{1/3}/r_d$~\footnote{$D_M(z)$ is defined as $D_M(z) = \frac{c}{H_0\sqrt{\Omega_k}} \, \sinh \left[\sqrt{\Omega_k}\int_0^z \frac{dz'}{H(z')/H_0}\right]$, and $D_H(z)$ is defined as $D_H(z) = c/H(z)$. The constraints on these ratios at different redshifts are detailed in Tab. 1 of~\cite{DESI:2024mwx}.}. 


\item[SN Ia] For a a complementary probe of the expansion history, we adopt type Ia supernovae (SN Ia) data from the PanthenonPlus compilation, which includes 1550 spectroscopically confirmed SN Ia in the redshift range $0.001 < z < 2.26$~\cite{Scolnic:2021amr}.
\end{description}

Combining these datasets and corresponding likelihoods, we perform MCMC analyses with the {\tt MontePython} engine~\cite{Audren:2012wb,Brinckmann:2018cvx} and the modified {\tt CLASS} code~\cite{Lesgourgues:2011re,Lesgourgues:2011rh}, incorporating state-of-the-art predictions for $Y_P$ and $D/H$ discussed above. In particular, we consider two approaches to the treatment of thermonuclear reaction rates in BBN: the PRIMAT-driven approach~\cite{Descouvemont:2004cw,Iliadis:2016vkw,InestaGomez:2017eya,deSouza:2018gdx,deSouza:2019pmr,Moscoso:2021xog,Pitrou:2018cgg,Pitrou:2020etk,Pitrou:2021vqr} and the NACRE II-driven approach~\cite{Xu:2013fha,Iliadis:2016vkw,Mossa:2020gjc,Fields:2019pfx}. While both approaches yield nearly identical predictions for $Y_P$, the PRIMAT-driven method predicts a significantly smaller $D/H$ due to differing nuclear rates for 10 key reactions in BBN~\cite{Pitrou:2018cgg,Burns:2022hkq,Li:2024gzf}.


For the MCMC run, we require convergence of all chains according to the Gelman-Rubin criterion, $R-1<0.01$, and use {\tt getDist}~\cite{Lewis:2019xzd} to derive the constraints presented in this letter. We employ two information criteria for model comparison: the change in chi-squared relative to the $\Lambda$CDM model, $\Delta \chi^2 = \chi^2_{\rm Model} - \chi^2_{\rm \Lambda CDM}$, and the Deviance Information Criterion (DIC)~\cite{Liddle:2007fy}, $\Delta DIC = \rm DIC_{\rm Model} - DIC_{\rm \Lambda CDM}$~\footnote{DIC is defined $DIC \equiv \rm <\chi^2> + p_D$, where $p_D$ is the effective number of parameters estimated from the posterior variance of the deviance, such that $p_D \equiv <\chi^2> - \chi^2(<\theta>)$.}. For both criteria, a lower value indicates a larger preference for the model over the $\Lambda$CDM model (see e.g., Table 1 of~\cite{Grandis:2016fwl}).



\begin{figure*}\centering
\includegraphics[width=0.4\textwidth]{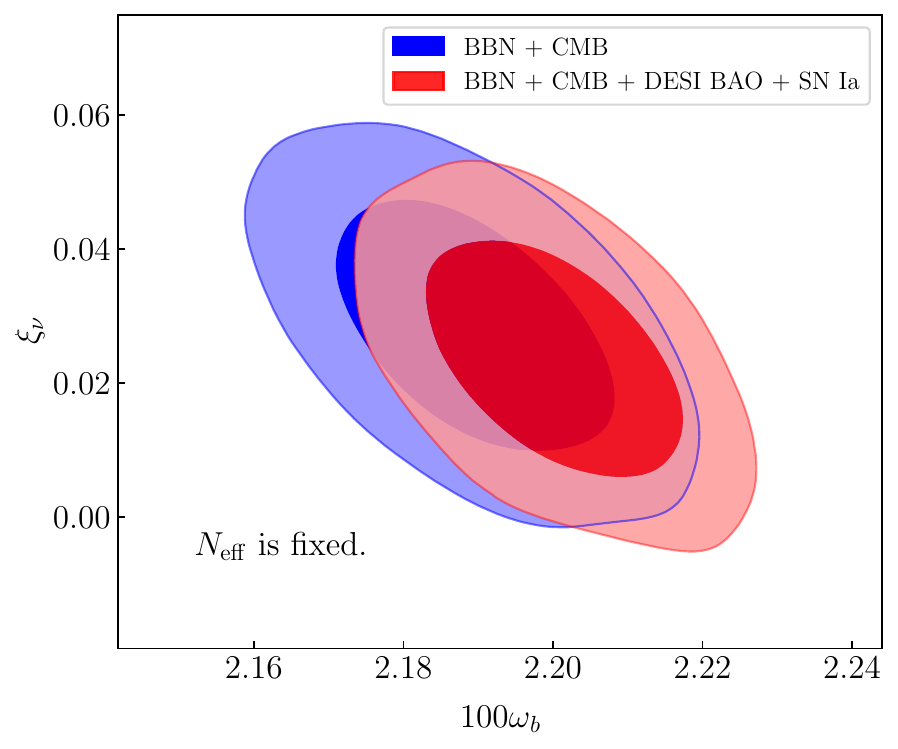} 
\includegraphics[width=0.4\textwidth]{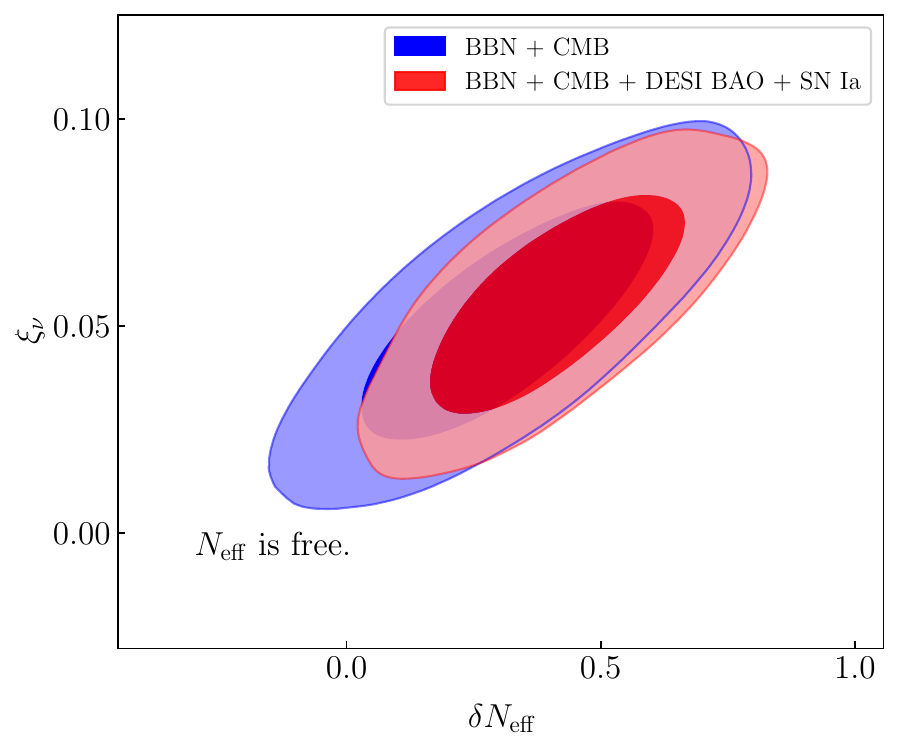} 
\caption{Marginalized posterior distributions of the baryon abundance $\omega_b$ and the neutrino degeneracy parameter $\xi_\nu$ at the $68\%$ and $95\%$ confidence levels, obtained from a state-of-the-art analysis using the PRIMAT-driven approach for BBN prediction. The blue and red regions correspond to results derived from two different datasets of cosmological observations.}
\label{fg:mcmc}
\end{figure*}

\begin{table*}[t]
\begin{center}
{\def\arraystretch{1.3}
\begin{tabular}{c|c|ccccc}
\hline\hline
\multicolumn{7}{c}{\textbf{Bounds for the model: $\mathbf{\Lambda {\rm \textbf{CDM}} + \sum m_\nu + \xi_\nu }$}}     \\ 
\hline \hline
 \textbf{Datasets}  & \textbf{BBN Approach}  & ${\bf \xi_{\nu}} $ & $\delta\neff $  & $ {\sum m_{\nu_i} \,\, (\ev) }$ & $100\times \omega_{b}$ &  $ H_0 $ [km/s/Mpc] \\
 \hline \hline
   \multirow{2}{*}{{\bf BBN + CMB}} &  {\bf PRIMAT}   & $0.028 \pm 0.012$   & --  &  $0.40 $ &  $2.190 \pm 0.012$ & $65.1^{+1.6}_{-0.9} $ \\ 
   \cline{2-7}
    &  {\bf NACRE II}   & $0.034 \pm 0.012 $ & --  & 0.30 & $2.209 \pm 0.012 $  & $66.1^{+1.2}_{-0.65}$ \\  
\hline \hline
    \multirow{2}{110 pt}{\bf BBN + CMB +DESI BAO + SN Ia} &  {\bf PRIMAT}   & $0.024 \pm 0.012$   & --  &  $0.14 $ &  $2.200 \pm 0.011$ & $66.99 \pm 0.42 $ \\ 
   \cline{2-7}
    &  {\bf NACRE II}   & $0.031 \pm 0.012 $ & --  & 0.14 & $2.215 \pm 0.011 $  & $ 67.23 \pm 0.45$ \\  
\hline \hline
\multicolumn{7}{c}{\textbf{Bounds for the model: $\mathbf{\Lambda {\rm \textbf{CDM}} + \sum m_\nu + \xi_\nu + \delta \neff}$}}     \\ 
\hline \hline
 \textbf{Datasets}  & \textbf{BBN Approach}  & ${\bf \xi_{\nu}} $ &  $\delta\neff $  & $ {\sum m_{\nu_i} \,\, (\ev) }$ & $100\times \omega_{b}$ &  $ H_0 $ [km/s/Mpc] \\
 \hline \hline
   \multirow{2}{*}{{\bf BBN + CMB}} &  {\bf PRIMAT}   & $0.051 \pm 0.018$   & $0.31 \pm 0.19 $ &  $0.34 $ &  $2.221 \pm 0.023$ & $67.7^{+2.1}_{-1.7} $ \\ 
   \cline{2-7}
    &  {\bf NACRE II}   & $0.041 \pm 0.019 $ & $0.08 \pm 0.19$  & 0.29 & $2.217 \pm 0.022 $  & $66.8^{+1.9}_{-1.6}$ \\  
\hline \hline
    \multirow{2}{110 pt}{\bf BBN + CMB +DESI BAO + SN Ia} &  {\bf PRIMAT}   & $0.056 \pm 0.017$   & $0.41 \pm 0.16$  &  $0.16 $ &  $2.236 \pm 0.018$ & $69.5 \pm 1.2 $ \\ 
   \cline{2-7}
    &  {\bf NACRE II}   & $0.048 \pm 0.018 $ & $0.20 \pm 0.17$  & 0.14 & $2.233 \pm 0.018 $  & $ 68.6 \pm 1.1$ \\  
\hline \hline
\end{tabular}
}
\end{center}
\vspace{-0.3cm}
\caption{
Summary of constraints for main parameters from considering several combinations of BBN, CMB, BAO and SN Ia data.  See main text for details.}
\label{tab:CurrentConstraints}
\end{table*}

\textbf{\textit{  Result A: State-of-the-art constraints on PNA.}}
The constraints on PNA derived from two cosmological datasets: the traditional "BBN + CMB" and the new "BBN + CMB + DESI BAO + SN Ia" data sets, which incorporate the LSS observations.

First, we consider the cosmological model $\Lambda {{\rm CDM}} + \sum m_\nu + \xi_\nu$, where $\neff$ is fixed to the value in Eq.~\eqref{eq:neff}. For both the PRIMAT-driven and NACRE II-driven BBN predictions, the corresponding $68\%$ bounds for $\omega_b$ and $\xi_\nu$ are shown in Table \ref{tab:CurrentConstraints}, while the PRIMAT-driven $68\%$ and $95\%$ marginalized posteriors of $\omega_b$ and $\xi_\nu$ are shown in the left panel of Figure \ref{fg:mcmc}.
For the "BBN + CMB" set, the constraints on $\xi_\nu$ are $\xi_\nu = 0.028 \pm 0.012$ for the PRIMAT driven approach and $\xi_\nu = 0.034 \pm 0.012$ for the NACRE II driven approach, with deviations of $2.3\sigma$ and $2.8\sigma$ from zero, respectively. 
Compared to previous results for the PRIMAT driven approach obtained in~\cite{Escudero:2022okz}, our results show a $\sim 18\%$ smaller mean value and a slight reduce on uncertainty.
With the LSS observation data, the resulting confidence region shifts towards a larger $\omega_b$, and hence a smaller $\xi_\nu$. Consequently, the bounds on $\xi_\nu$ become:
\begin{align}
    \xi_\nu &= 0.024 \pm 0.012 \quad &[\rm PRIMAT \,\, driven], \\
    \xi_\nu &= 0.031 \pm 0.012 \quad &[\rm NACRE \,\, II \,\, driven],
\end{align}
where the significance of the deviations is reduced to $2.0\sigma$ and $2.6\sigma$, respectively.

Allowing $\neff$ to vary freely significantly alters the constraints on $\xi_{\nu}$ and other parameters. As before, the resulting $68\%$ confidence level constraints are summarized in Table \ref{tab:CurrentConstraints}, while the PRIMAT-driven $68\%$ and $95\%$ confidence regions for $\delta \neff$ and $\xi_\nu$ are shown in the right panel of Figure \ref{fg:mcmc}.
In general, allowing $\neff$ to vary freely increases both the central value and the uncertainties for the constraints on $\xi_\nu$, while the constraints on $\delta \neff$ show significant variation depending on the assumed nuclear rates for BBN.
For the "BBN + CMB" sets, the PRIMAT-driven approach yields a positive $\delta \neff = 0.31 \pm 0.19$, whereas the NACRE II-driven approach gives $\delta \neff = 0.08 \pm 0.19$, showing no significant preference~\footnote{Compared to the results obtained in ~\cite{Escudero:2022okz} with the PRIMAT approach, our analysis gives an almost identical central value with a slight reduction of the uncertainty for $\xi_\nu$, while the preference for positive $\delta \neff$ is more significant in our analysis.}.
More importantly, including the LSS observations in this analysis shifts the confidence region toward both a larger $\xin$ and a larger $\delta \neff$. As a result, the constraints on $\xi_\nu$ and $\delta \neff$ for the NACRE II-driven approach are:
\begin{equation}
    \xi_\nu = 0.048 \pm 0.018 \, , \quad\delta \neff = 0.20 \pm 0.17
\end{equation}
indicating a preference for a non-zero $\xi_\nu$ at the $2.7 \sigma$ level and a non-zero $\delta \neff$ at the $1.2 \sigma$ level. For the PRIMAT-driven approach, the constraints become:
\begin{equation}
    \xi_\nu = 0.056 \pm 0.017 \, , \quad \delta\neff = 0.41 \pm 0.16
\end{equation}
with a preference for a non-zero $\xi_\nu$ at the $3.3 \sigma$ level and a non-zero $\delta\neff$ at the $2.6 \sigma$ level~\footnote{It is worth noting that the required positive $\delta \neff$ cannot be fully explained by the contribution of $\xi_\nu$ under the assumption of flavor equilibration ($\xi_{\nu_e} = \xi_{\nu_\mu} = \xi_{\nu_\tau} = \xi_\nu$). However, it may be interpreted as the result of a large neutrino asymmetry in the muon-tau sector in the case where flavor equilibration is not fully realized, as discussed in the recent study~\cite{Froustey:2024mgf}.}. 
This preference for a non-zero $\delta \neff$ in both the BBN NACRE/PRIMAT-driven approaches arises from the need to fit the DESI BAO observations, and it also help alleviate the $H_0$ tension, as we will discuss later.

\begin{figure}\centering
\includegraphics[width=0.4\textwidth]{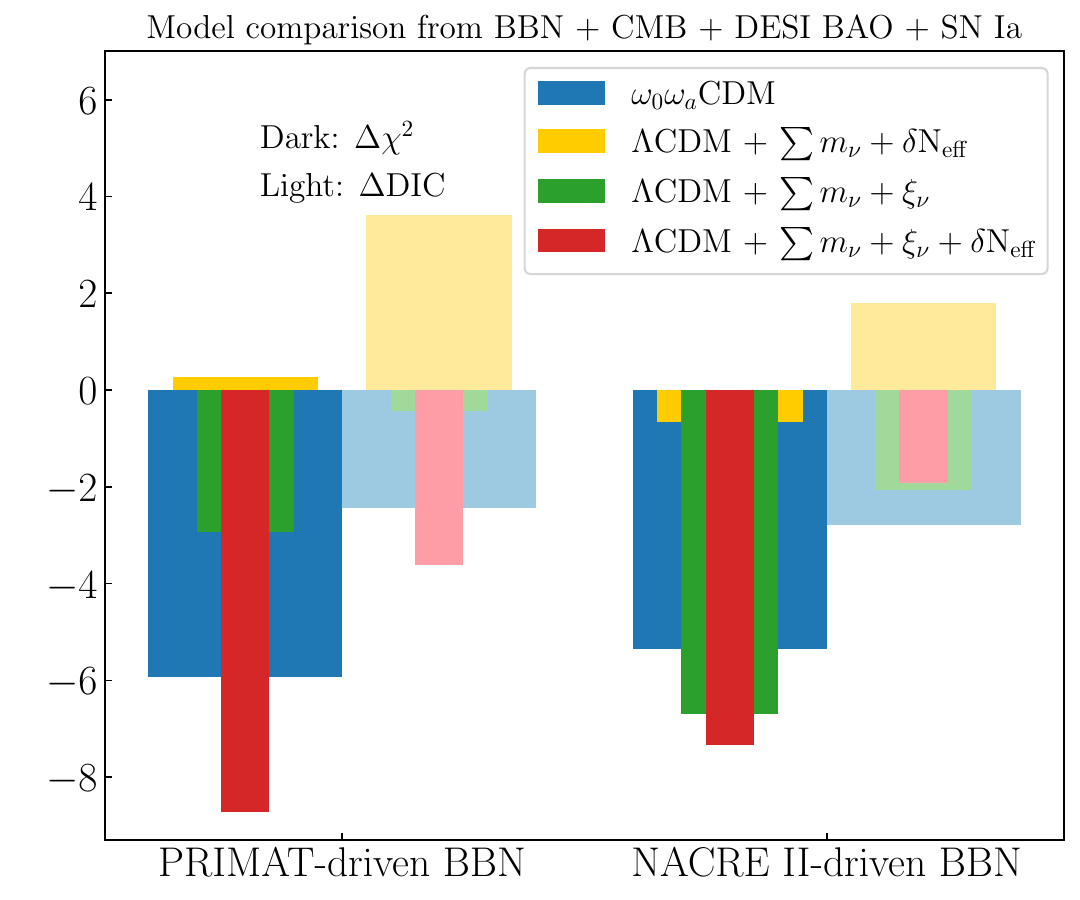} 
\caption{
 Values of two information criteria, $\Delta DIC$ and $\Delta \chi^2$ for various models including the $\omega_0 \omega_a$CDM and neutrinophilic $\Lambda$CDM extensions. 
 A lower value indicate a larger preference for the model over the $\Lambda$CDM model.
}
\label{fg:model comparison}
\end{figure}


\textbf{\textit{  Result B: Model Comparison with EMPRESS and DESI.}} Using $\Lambda$CDM model as a benchmark, we will consider $\omega_0 \omega_a$CDM model and various neutrinophilic $\Lambda$CDM extensions with $\xi_\nu$ or (and) $\delta \neff$ based on data from BBN, CMB, DESI BAO, and SN Ia. Values of the two information criteria we considered, $\Delta\chi^2$ and $\Delta DIC$, are shown in Figure \ref{fg:model comparison} with dark and light bars, respectively. 
For the $\omega_0 \omega_a$CDM model, both the PRIMAT-driven and NACRE II-driven approaches yield $\Delta\chi^2 \approx -6$ and $\Delta DIC \approx -2$, indicating a mild preference over $\Lambda$CDM, which is consistent with the finding in~\cite{DESI:2024mwx}.
For the $\Lambda$CDM + $\sum m_\nu$ + $\xi_\nu$ model, we see that the results depend on the BBN NACRE/PRIMAT-driven approaches, where the model performs better or similarly than the $\omega_0 \omega_a$CDM model based on the NACRE II-driven approach, while it performs worse based on the PRIMAT-driven approach.
For the $\Lambda$CDM + $\sum m_\nu$ + $\delta \neff$ model, $\Delta\chi^2$ and $\Delta DIC$ are positive or nearly zero, showing an even worse performance than the $\Lambda$CDM model in both BBN approaches.  
Finally, we see that the $\Lambda$CDM + $\sum m_\nu$ + $\xi_\nu$ + $\delta \neff$ model yields a similar or smaller $\Delta\chi^2$ and $\Delta DIC$ in both BBN approaches, indicating a stronger preference over the $\omega_0 \omega_a$CDM~\footnote{For the NACRE II-driven approach, the $\Delta DIC$ criterion shows a less preference of the $\Lambda$CDM + $\sum m_\nu$ + $\xi_\nu$ + $\delta \neff$ model than the $\omega_0 \omega_a$CDM model, which is mainly caused by the fact that $\Delta DIC$ criterion particularly favors models with much less parameters. }.

\begin{figure}\centering
\includegraphics[width=0.4\textwidth]{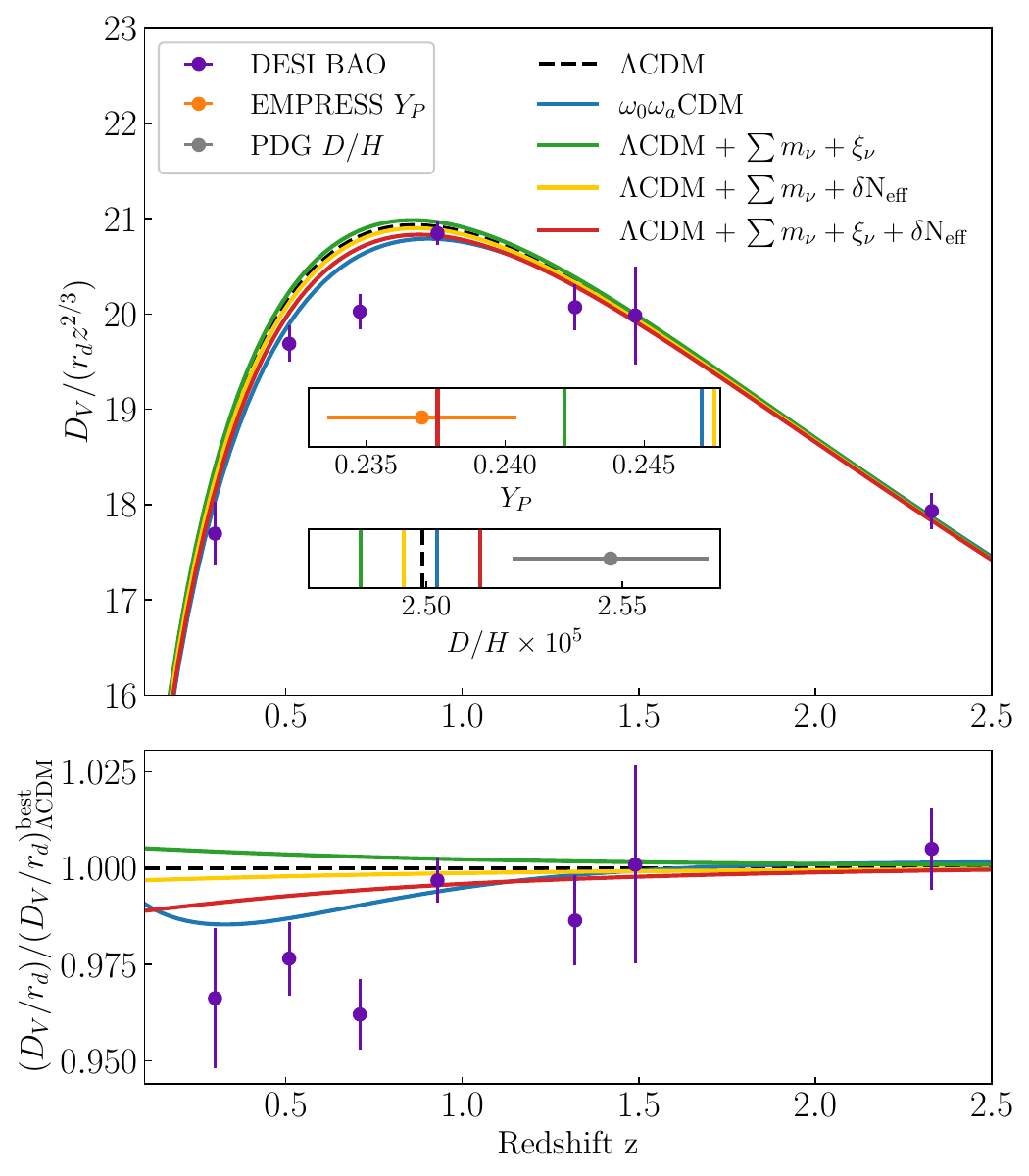} 
\caption{
Top panel: The DESI BAO results of the angle-averaged distance $D_V/r_d$ normalized by $z^{-2/3}$ at various redshifts. The dashed and solid lines represent best-fit predictions from the $\Lambda$CDM model, $\omega_0 \omega_a$CDM model, and various neutrinophilic $\Lambda$CDM extensions. The measurements of $Y_P$ and $D/H$ with the corresponding best-fit predictions are also included. 
Bottom panel: The same data points for DESI BAO and models as in the top panel, but normalized by the best-fit $\Lambda$CDM predictions.
}
\label{fg:bestfit}
\end{figure}

To better understand the comparison results for these models, we show in Figure~\ref{fg:bestfit} the BBN and DESI BAO measurements, along with corresponding best-fit predictions from the $\Lambda$CDM, $\omega_0 \omega_a$CDM, and various neutrinophilic $\Lambda$CDM extensions, based on the PRIMAT-driven approach for the BBN predictions.
In the upper panel of Fig.~\ref{fg:bestfit}, the DESI BAO measurements of the angle-averaged distance $D_V/r_d$ at different redshifts are normalized by $z^{-2/3}$, while in the lower panel, they are normalized with the best-fit $\Lambda$CDM predictions. The BBN measurements and corresponding best-fit predictions are also included in the upper panel.
Concerning the BBN measurements, we see that only the $\Lambda$CDM + $\sum m_\nu + \xi_\nu + \delta \neff$ model can fit $Y_P$ well while the $D/H$ result roughly fits. The best-fits of all the other models cannot fit $Y_P$ and $D/H$ at the same time, which actually shows a slight tension between current $Y_P$ and $D/H$ measurements, see~\cite{Li:2024gzf} for relevant discussions.
Concerning the DESI BAO measurements, we find that while $\omega_0 \omega_a$CDM can fit the smaller $D_V/r_d$ at low redshifts due to the modified late-time expansion, the $\Lambda$CDM + $\sum m_\nu + \xi_\nu + \delta \neff$ model can also provide a satisfactory fit, which is caused by the positive $\delta \neff$ contributions needed to fit BBN measurements. 
In summary, only the $\Lambda$CDM + $\sum m_\nu + \xi_\nu + \delta \neff$ model can fit both the BBN and DESI BAO observations well, which makes it the most preferred model.

Next we briefly discuss the results in the NACRE II-driven approach. Since the NACRE II-driven approach predicts a smaller $D/H$, both neutrinophilic $\Lambda$CDM extensions with PNA can fit $Y_P$ and $D/H$ measurements well, leading to the preference of the $\Lambda$CDM + $\sum m_\nu$ + $\xi_\nu$ model than the $\omega_0 \omega_a$CDM model in Figure~\ref{fg:model comparison}.
On the other hand, the best-fit of the $\Lambda {\rm CDM} + \sum m_\nu + \xi_\nu + \delta \neff$ model for the DESI BAO results becomes worse, as the late-time expansion history of the universe is less affected with the small $\delta \neff$ contribution in this case.


\begin{figure}\centering
\includegraphics[width=0.4\textwidth]{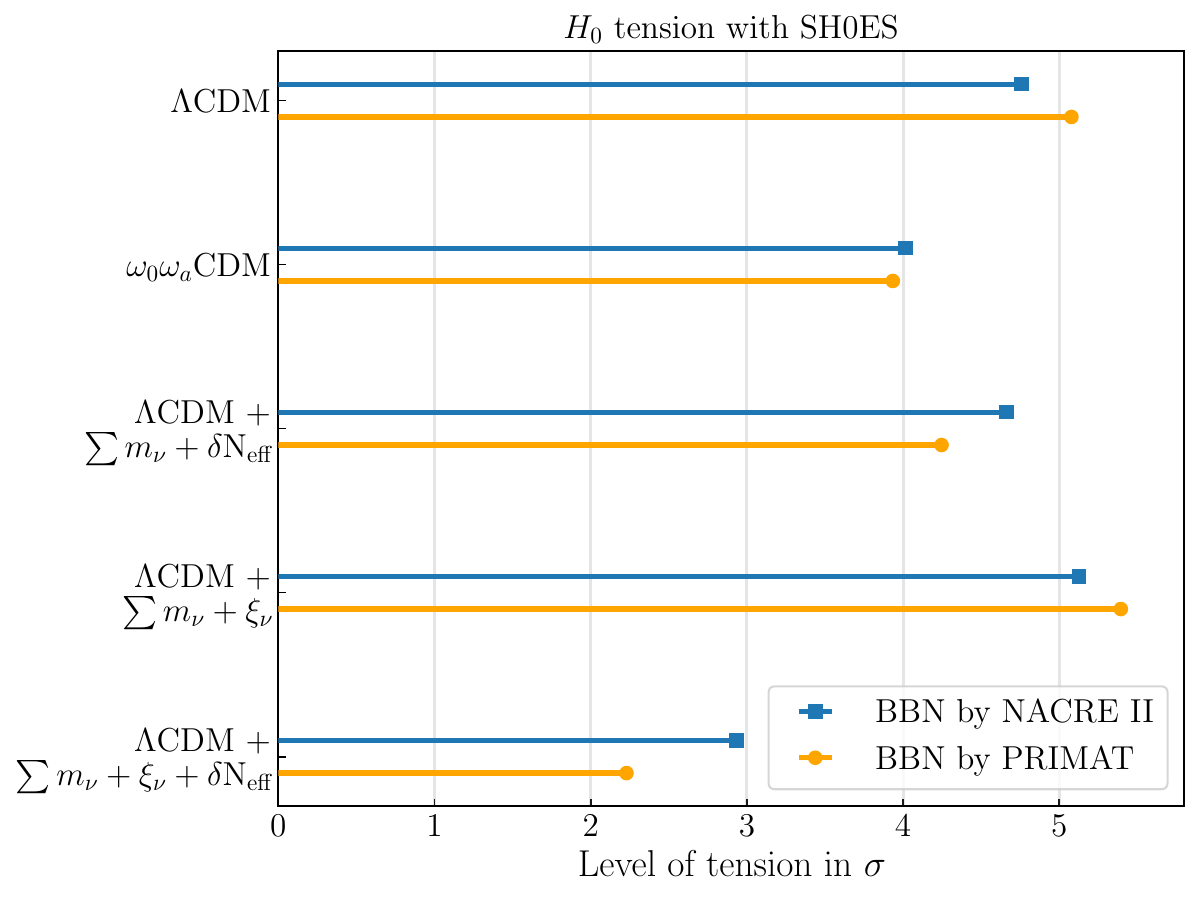} 
\caption{A summary of the Hubble tension obtained from the combination of BBN, CMB, DESI BAO and SN Ia results and the SH0ES result of~\cite{Riess:2021jrx}, assuming various cosmological models and approaches for BBN predictions.
}
\label{fg:H0 tension}
\end{figure}

\textbf{\textit{  Result C: Alleviate the Hubble tension.}} For various cosmological models we considered, the results of Hubble parameter are summarized in Table~\ref{tab:CurrentConstraints}, while the significance of the Hubble tension compared to the SH0ES result~\cite{Riess:2021jrx} for both BBN nuclear approaches are shown in Figure \ref{fg:H0 tension}. 
For both the $\Lambda$CDM model and its extensions with $\xi_\nu$ or $\delta \neff$, the Hubble tension remains at $\sim 5\sigma$, while the $\omega_0 \omega_a$CDM model reduces it to $\sim 4\sigma$. However, for the $\Lambda {\rm CDM} + \sum m_\nu + \xi_\nu + \delta \neff$ model, the central value and uncertainties for the Hubble parameter both increase from the contribution of additional $\delta \neff$. Consequently, Hubble tension is reduced to $2.2 \sigma$ for the PRIMAT approach and $2.9\sigma$ for the NACRE II approach.


\textbf{\textit{  Summary and Outlook.}} Our study presents the first detailed analysis that combines attractive indications from EMPRESS BBN, DESI BAO measurements, and Hubble tension. While these are traditionally studied in detail individually, we show that a simple $\Lambda {\rm CDM} + \sum m_\nu + \xi_\nu + \delta \neff$ model can explain both the EMPRESS BBN and DESI BAO results, while reducing the Hubble tension to the $2.2 \sigma$ level at the same time.


Along this direction, there are still many things to be discussed. 
First, we consider the $S_8$ tension, where $S_8 = \sigma_8 \sqrt{\Omega_8/0.3}$, and $\sigma_8$ represents the amplitude of matter fluctuations on $8 h^{-1}$ scales. Similar to the Hubble tension, the $S_8$ tension manifests as a $2-3\sigma$ discrepancy between early universe measurements and late universe cosmic shear data~\cite{DES_2021,DiValentino:2020vvd,DiValentino:2018gcu,Kilo-DegreeSurvey:2023gfr,Troster:2019ean,Heymans:2020gsg,Dalal:2023olq,ACT:2024okh,DES:2024oud}. For our $\Lambda {\rm CDM} + \sum m_\nu + \xi_\nu + \delta \neff$ model, the MCMC analysis gives $S_8 = 0.840 \pm 0.011$ (PRIMAT-driven BBN) and $S_8 = 0.834 \pm 0.011$ (NACRE II-driven BBN), also showing a $3.0\sigma$ and $2.8\sigma$ tension with the DES-Y3 measurement~\cite{DES:2021bvc}, respectively.  


We also note that the neutrino properties constraints are sensitive to the BBN observations. For instance, replacing the EMPRESS $Y_P$ observation with the PDG value $Y_P = 0.2475 \pm 0.003$~\cite{pdg} removes the preference for non-zero $\xin$ and $\delta \neff$. As a result, the $\Lambda {\rm CDM} + \sum m_\nu + \xi_\nu + \delta \neff$ model no longer accommodates DESI BAO data or the Hubble tension. This highlights the critical role of accurate BBN measurements, which will remain an important probe of new physics in the future.

\begin{acknowledgements}

\textbf{\textit{Acknowledgements.}} This work is supported by the National Science Foundation of China under Grants No. 12347105, No. 12375099 and No. 12047503, and the National Key Research and Development Program of China Grant No. 2020YFC2201501, No. 2021YFA0718304. The authors gratefully acknowledge the use of publicly available codes {\tt FortEPiaNO}~\cite{Gariazzo:2019gyi,Bennett:2020zkv}, {\tt PRIMAT}~\cite{Pitrou:2018cgg}, {\tt CLASS}~\cite{Lesgourgues:2011re,Lesgourgues:2011rh}, {\tt MontePython}\cite{Audren:2012wb,Brinckmann:2018cvx} and {\tt getDist}~\cite{Lewis:2019xzd}.

\end{acknowledgements}

\bibliographystyle{apsrev4-1}
\bibliography{references}

\end{document}